\documentstyle[12pt]{article}

\begin{document}

\title{SCALAR FIELD COSMOLOGIES WITH \\
PERFECT FLUID IN ROBERTSON-WALKER METRIC}

\author{Luis P. Chimento and Alejandro S. Jakubi      \\
{\it Departamento de F\'{\i}sica,  }\\
{\it  Facultad de Ciencias Exactas y Naturales, }\\
{\it Universidad de Buenos Aires }\\
{\it  Ciudad  Universitaria,  Pabell\'{o}n  I, }\\
{\it 1428 Buenos Aires, Argentina.}}

\maketitle

\begin{abstract}
Several isotropic, homogeneous cosmological models containing a
self-interacting minimally coupled scalar field, a perfect fluid source and
cosmological constant are solved. New exact, asymptotically stable solutions
with an inflationary regime or a final Friedmann stage  are found for some
simple, interesting potentials. It is shown that the fluid and the curvature
may determine how these models evolve for large times.

\end{abstract}

\vskip 3cm


\newpage

\section{Introduction}

A self-interacting scalar field has been introduced in cosmological models as
a matter source to the Einstein equations because, when dominated by the
potential energy, it violates the strong energy condition and drives the
universe into an inflationary period \cite{Gut},\cite{Alb},\cite{Lin82}.
Scalar fields also play a fundamental role in unified theories of the strong,
weak and electromagnetic interactions through the mechanism of spontaneous
symmetry breaking. Of particular interest for cosmology are those symmetries
which spontaneously broken today were restored in the hot early universe.
Thus, several kinds of topological defects may have been produced
\cite{Lin},\cite{Kol}. Further, a scalar field has been proposed as a kind of
dark matter \cite{Rat}.

Currently, there is no underlying principle that uniquely specifies the
potential for the scalar field and many proposals have been considered. Some
were based in new particle physics and gravitational theories
\cite{Lin}.
Others were postulated ad-hoc to obtain the desired evolution. For instance,
Ellis et al. \cite{Ell} have proposed a scheme to find the potential
function for a given expansion of a Robertson-Walker universe. Also, a
formalism has been proposed to reconstruct the potential from knowledge of
tensor gravitational spectrum or the scalar density fluctuation spectrum
\cite{Cop}, \cite{Lid94}.

Notwithstanding the number of papers devoted to understand the dynamics of
the
self-interacting scalar field in curved spacetimes, very little is known yet
about exact solutions of these cosmological models. Most of these solutions
correspond to models in a spatially flat Robertson-Walker metric which have no
other source besides the scalar field \cite{Gut}, \cite{Luc}, \cite{Bur},
\cite{Rit1}, \cite{Rit2}, \cite{Mus}, \cite{Sal}, \cite{Bar90}, \cite{Chi},
\cite{Bar94}, \cite{Mie}.
Few exact solutions are known with spatial curvature \cite{Hal}, \cite {Ell},
\cite{Oze}, \cite{Eas}, or a perfect fluid \cite{Rat}, \cite{Bar93},
\cite{Cos};  and we are aware of none with both terms. The problem arises in
the non-linearity of the system of differential equations for the scalar and
gravitational fields. Usually, in studies of inflationary dynamics, the
evolution is divided in several steps, and during each of them some kind of
approximation is assumed to simplify the system of equations
\cite{Lin}, \cite{Kol}.

In section 2 of this paper we show a procedure to reduce to quadratures the
Einstein-scalar field equations in a Robertson-Walker metric
for an arbitrary potential, a perfect baryotropic fluid
source and a cosmological constant. This allows us to address the issue of
whether the scalar field is always the dominating force in driving the
evolution of the
universe or the fluid and curvature terms may also play a role.
We analyse in section 3 general properties of the solutions and we study their
asymptotic stability by means of the method of Lyapunov \cite{Kra}. We
consider also their structural stability. Our procedure is illustrated in
section 4 with several examples. First, we use it to rederive and generalize a
few solutions for the well known exponential potential, and then we present
some new exact solutions for several simple, interesting potentials. Some of
them are suitable for cosmological models based on the chaotic inflationary
scenario. No a priori assumption like a slowly varying field is required to
perform the calculations, so that we can check the validity of this
assumption.
Finally, the conclusions are stated in section 5.

\section{The Model}

We wish to investigate the evolution of a universe filed with a scalar field
and a perfect fluid. The scalar field $\phi $ has a self-interaction
potential $V(\phi)$ and is minimally coupled to gravity, so that it obeys
the Klein-Gordon equation

\begin{equation}
\label{1}
\Box \phi +{\frac{dV}{d\phi }}=0
\end{equation}
The perfect fluid has four-velocity $u_i$ and its pressure $p$ and energy
density $\rho $ are related by the equation of state

\begin{equation}
\label{2}
p = ( \gamma -1 ) \rho
\end{equation}
\noindent with a constant adiabatic index $0\le \gamma \le 2$. Thus, we must
solve equation (\ref{1}) together with the Einstein equations

\begin{equation}
\label{3}
R_{ik}-\frac 12g_{ik}R+\Lambda g_{ik}=T_{ik}^\phi +T_{ik}^f
\end{equation}
We are using units such that $c=8\pi G=1$, $\Lambda $ is the cosmological
constant and

\begin{equation}
\label{4a}
T_{ik}^\phi =\phi _{;i}\phi _{;k}-g_{ik}\left[ {\frac 12}\phi
_{;m}\phi ^{;m}-V(\phi )\right]
\end{equation}
\begin{equation}
\label{4b}
T_{ik}^f=(\rho +p)u_iu_k-pg_{ik}
\end{equation}

\noindent
are the stress-energy tensors of the field and the fluid.

In a Robertson-Walker metric

\begin{equation}
\label{5}
ds^2=dt^2-a^2(t)\left [{\frac{dr^2}{1-kr^2}}+r^2(d\theta ^2+\sin
{}^2\theta d\phi ^2)\right]
\end{equation}

\noindent with scale factor $a(t)$ and curvature parameter $k=0,\pm 1$,
 equations (\ref{1}) and (\ref{3}) become

\begin{equation}
\label{6}
\ddot \phi +3H\dot \phi +{\frac{dV}{d\phi }}=0
\end{equation}
\begin{equation}
\label{7}
3H^2=\frac {1}{2}{\dot \phi }^2+V(\phi )+\rho -3{\frac k{a^2}}+\Lambda
\end{equation}

\noindent where the dot means $d/dt$, $H=\dot a/a$ and $\phi =\phi (t)$. Also,
from the conservation of (\ref{4b}), $\rho
=\rho _0/a^{3\gamma }$ where $\rho _0\ge 0$ is a constant.

To  integrate this system we make the change of
variables $dt=a^3 d\eta$ in (\ref{6})

\begin{equation} \label{71}
\frac{d^2\phi}{d\eta^2}+a^6 \frac{dV}{d\phi}=0
\end{equation}

\noindent
and we write

\begin{equation}
\label{10}
V[\phi (a)] = {\frac{F(a)}{a^{6}}}
\end{equation}

\noindent
Thus we obtain the first integral of (\ref{6})

\begin{equation}
\label{11}
\frac {1}{2}{\dot \phi }^2+V(\phi)-{\frac 6{a^6}\ \int }da{\frac F a}={
\frac C{a^6}}
\end{equation}

\noindent where $C$ is an arbitrary integration constant. Then, using
the scale factor as the independent variable and
equations
(\ref{7}) and (\ref{11}), we have reduced the problem to quadratures:

\begin{equation}
\Delta t = {\sqrt{3}\int \frac{da}{a}\left[
{\frac 6{a^6}\int }da{\frac Fa}+{\frac C{a^6}}+\rho -3{
\frac k{a^2}}+\Lambda\right]^{-1/2}}
\label{12}
\end{equation}

\begin{equation}
\Delta \phi = {\sqrt{6}\int \frac{da}a\left[\frac{-F+6\int da F/a+C}
{6\int da F/a+C+\rho a^6-3ka^4+\Lambda a^6}\right]}^{1/2}
\label{13}
\end{equation}

\noindent where $\Delta t\equiv t-t_0$, $\Delta \phi \equiv \phi -\phi _0$
and $t_0$, $\phi _0$ are two other arbitrary integration constants.

Equations (\ref{10}), (\ref{12}) and (\ref{13}) allows us to
obtain a solution $a(t), \phi(t)$ for the potential $V(\phi)$, given a
function $F(a)$, the parameters $\rho_0$, $\gamma$, $k$ and $\Lambda$, and a
choice of the primitive $\int da F/a$ and the integration constant $C$.

\section{ General Behavior of Solutions}

In general the set of solutions that arise from (\ref{12}) (\ref{13})
contains complex solutions without physical sense that must be rejected. So,
to obtain physical solutions we require that the functions

\begin{equation}
G(a)= {\frac 6{a^6}\int }da{\frac Fa}+{\frac C{a^6}}+\rho -3{
\frac k{a^2}}+\Lambda
\label{15}
\end{equation}

\begin{equation} \label{152}
L(a)= -\frac {F}{a^6}+\frac {6}{a^6}\int da
\frac {F}{a}+\frac {C}{a^6}
\end{equation}

\noindent
are nonnegative on an interval $(a_1,a_2)$,
$0\le a_1<a_2\le \infty$. Since $G(a)=3H^2$ and $L(a)=\dot\phi^2/2$, a
sufficient condition is that $a>0$, $\dot a\neq0$ and $\dot\phi\neq 0$. To
relax these conditions we need to
study the behavior of the scale factor, which is
determined by the function $G(a)$. Thus, a
solution $a(t)$ is monotonic ($\dot a\neq 0$) when $G(a)>0$ but it is bounded
if $G(a_0)=0$ for $0<a_0<\infty$. A growing monotonic solution starts from
$a=0$ and expands without bound. A solution is singular when $a(t)$ vanishes
at a finite time, that is when the integral

\begin{equation}
T\left(a_1,a_2\right)=\int_{a_1}^{a_2}\frac{da}{a\sqrt{G(a)}}
\label{155}
\end{equation}

\noindent
converges in the limit $a_1\to 0$. Let us assume that in this limit the
leading behaviors of $F(a)$ and $\int da F/a$ are $ \bar K a^n$ and $\bar K
a^n/n$, where $\bar K>0$ and $n\neq 0$ are constants. So $G(a)\sim K/a^\alpha
$ where $K>0$ is another constant and $\alpha$ is the maximum of the exponents
in (\ref{15}), namely $(6-n,6,3\gamma,2)$. Then solutions are singular when
$\alpha>0$, which occurs when any of the following conditions are met: $C>0$,
$\rho_0>0$, $k=-1$ or $n<6$. Singular solutions have particle horizons when
$\alpha >2$, that is when  $C>0$ or $n<4$ or $\gamma>2/3$. Similarly, assuming
$F(a)\sim a^n$ for $a\to\infty$, we find that no solution diverges at a finite
time.

For bounded solutions,
we observe that the scale factor reaches $a_0$ at a finite time $t_0$ when
$T(a_1,a_2)$ converges in the limit $a_1\to a_0$. Let us assume
that  $G(a)\sim K(a^\delta-a_0^\delta)^\beta$, $\beta>0$ and $\delta\neq 0$
constants, in a neighborhood of $a_0$. If
$\beta<2$ then $a_0$ is a local extremum of $a(t)$ and the evolution is
time-symmetric about $t_0$. Otherwise, for $\beta\ge 2$, $a(t)\to a_0$ when
$|t|\to\infty$.

Provided that the requirements for a physical solution are met, the freedom to
select $F(a)$ reflects the freedom to choose the potential. However, only in
restricted cases the functions $a(t)$, $\phi(t)$ and $V(\phi)$ arise in closed
explicit form.

Though the solution (\ref{12}) (\ref{13}) of the system of equations
(\ref{6}) (\ref{7}) has three constants of integration, it does not provide
the general solution for a given potential. Rather it provides a special
solution for each member of a family of potentials. So, for each potential we
need to study the stability of this solution.

For open or flat models such that $V(\phi)$ has a local minimum at $\phi _m$
and $\Lambda_{eff}\equiv V(\phi _m)+\Lambda \ge 0$, we can study the stability
of solutions with asymptotic behavior $\phi (t)\rightarrow \phi _m$. First we
note that $\phi=\phi_m$ and $a(t)$ given by

\begin{equation}
\Delta t=\sqrt{3}\int \frac{da}{a}\left( \rho -\frac{3k}{a^2}+\Lambda
_{eff}\right) ^{-1/2}
\label{150}
\end{equation}

\noindent
is a classical solution of (\ref{6}) and (\ref{7}) with the effective
cosmological $\Lambda_{eff}>0$  \cite{Har} ($a(t)$ may not be given by
(\ref{150}) when $\Lambda_{eff}=0$).

To study the asymptotic stability of a solution it becomes convenient to
turn it into a fixed point. For the models under consideration $a(t)$ is
monotonic we may use $a$ as the independent variable. Then we can restrict the
analysis to the phase space $(\phi,\phi')$ and study the stability
of the point $(\phi_m,0)$. We find that the energy density of the field

\begin{equation}
\rho_\phi=V(\phi)+\frac{1}{2}{{\dot \phi }^2}
\label{151}
\end{equation}

\noindent
is a suitable a Lyapunov function as
it satisfies $\rho_\phi(\phi,\phi',a)>-\Lambda$ and ${\rho_\phi}'=-3aH^2
\phi'^2<0$ in a neighborhood of $(\phi _m,0)$. So this point is an attractor
and any
solution such that $\phi \rightarrow \phi _m$ for $a\rightarrow \infty $
(equivalently $t\rightarrow \infty $) is asymptotically stable. In this limit,
when $\Lambda_{eff}>0$, equation (\ref{150}) gives the leading behavior of
$a(t)$. Since

\begin{equation}
\dot H=-{\frac 12\dot \phi }^2-{\frac 12}\gamma \rho +{\frac k{a^2}}
\label{14}
\end{equation}

\noindent
is negative we note  that these solutions have a deflationary behavior.

On the other hand, in closed models some solutions are non-monotonic. However,
for any solution such that the scale factor is monotonic starting from a given
time, we may  apply the same analysis as before about its stability. Also, for
$k=1$ we have the scalar field analog of the Einstein Universe \cite{Har}. In
effect using (\ref{6}), (\ref{7}) and (\ref{14}) we find for a nonconstant
potential a unique static solution $a=a_E>0$ and $\phi=\phi_E$, provided that
$V'(\phi_E)=0$,

\begin{equation} \label{141}
V(\phi_E)=\frac{3\gamma-2}{\gamma a_E^2}-\Lambda
\end{equation}

\noindent
and $\gamma \rho_0=2a_E^{3\gamma-2}$.  In the case $\gamma=2/3$ there is no
particular equilibrium density and there is a static solution for any $a_E$.
To study the stability of the static solution we consider small perturbations
about it. In the linear regime the system decouples and we find that the
Einstein solution is stable when $\gamma< 2/3$ and $V''(\phi_E)>0$, and it is
unstable otherwise.

Recently it has been considered the question whether small variations in the
form  of the potential induce small variations in the scale factor and vice
versa. A model with this property is called structurally stable or "rigid"
\cite{Ell91b}. In the case that $k=\Lambda=\rho=0$, the rigidity of
inflationary models has been shown \cite{Lid93}. We wish to investigate
whether this property is also valid without these restrictions. Let us
consider a family of functions  $F(a)$ labeled by a parameter $r$. If $F(a,r)$
is a continuous function and $a>0$,  using (\ref{10}) and (\ref{13}),  a small
change $\delta  r$ induces a small change $\delta V$ in $V(\phi)$. Also, using
(\ref{12}), it induces a small change $\delta a$ in $a(t)$. Clearly, small
neighbourhoods in the space of scale factors and in the space of potentials
map continuously into each other. Thus we conclude that models where $F$ is
continuous are structurally stable.

\section{ Examples}

In this section we obtain
several new exact solutions for simple interesting potentials and
we concentrate mainly on models with spacial curvature and a fluid source
since very few exact solutions are known for them.

\subsection{Exponential and exponential-like potentials}

We consider first potentials of the form

\begin{equation}\label{181}
V(\phi)=B \exp(-\sigma \Delta\phi)
\end{equation}

Exponential potentials like (\ref{181}) have been extensively considered by
many authors, mainly in the context of "power-law" inflationary models. We
rederive and generalize exact solutions for this potential. For this we take

\begin{equation}
F(a) = B a^{s}
\label{16}
\end{equation}

\noindent
where $B$ and $s$ are constants, we choose $\int da F/a=Ba^s/s$ with
$C=0$ and we obtain a real solution when $0<s\le 6$  for $B>0$. Since $s=6$
corresponds to a constant potential, we consider this case no further. For
$s<6$, the energy density of the scalar field redshifts as a perfect fluid
with equation of state $p_\phi=(\gamma_\phi-1)\rho_\phi$:
$\rho_\phi=\rho_{0\phi}/a^n$, where $\rho_{0\phi}=6B/s$ and
$n=3\gamma_\phi=6-s$ \cite{Rat}. Using (\ref{11}) and (\ref{12}) we find a
power-law solution of the form

\begin{equation} \label{182}
a(t)=\left(K\Delta t\right)^\lambda\qquad
\Delta\phi=M\ln a
\end{equation}
\noindent
with some constants $K$, $M$ and $\lambda$ which are functions of the
parameters of the system.

\bigskip\noindent
Case 1. $\rho _{0}=k=\Lambda=0$.

$$
\sigma=(6-s)^{1/2}\qquad
K=\left(\frac{B}{2s}\right)^{1/2}(6-s)\qquad
\lambda=\frac{2}{6-s}\qquad
M=(6-s)^{1/2}
$$

\bigskip\noindent
This solution has been found by Lucchin et al. \cite{Luc}, its asymptotic
stability has been analysed by Halliwell \cite{Hal} and the general solution
for this model has been found by Salopek et al. \cite{Sal}.

\noindent
Case 2. $\rho=\Lambda=0$, $s=4$, $k=0,-1$ or $k=1$ with $B>2$.

$$
\sigma=\left[\frac{2}{B}(B-2k)\right]^{1/2}\qquad
K=\left(\frac{B}{2}-k\right)^{1/2}
$$
$$
\lambda=1\qquad
M=\left(\frac{2B}{B-2k}\right)^{1/2}
$$

\noindent
This solution has been found by Ellis et al. \cite{Ell} (see also \cite{Hal}
for $k=-1$).

\bigskip\noindent
Case 3. $k=\Lambda=0$, $s=6-3\gamma$.

$$
\sigma=\left[(6-s)\left(1+\frac{\rho_0 s}{6B}\right)\right]^{1/2}\qquad
K=\left(\frac{6B+s\rho_0}{3s}\right)^{1/2} \frac{6-s}{2}
$$
$$
\lambda=\frac{2}{6-s}=\frac{2}{3\gamma}\qquad
M=\left[\frac{6B(6-s)}{6B+\rho_0 s}\right]^{1/2}
$$

This solution has been found by Ratra et. al \cite{Rat} in the case
$\gamma=1$, and we extend it here to arbitrary $\gamma$. It generalizes case 1
for a fluid source and for $\gamma>2/3$ it is not inflationary and has
particle horizons. It is usually assumed that an exponential potential with
$\sigma$ small enough leads to a power-law inflationary evolution
($\lambda>1$), so that any matter term becomes rapidly dominated by the scalar
field energy density. However in our exact solution, which is  asymptotically
stable, the energy densities of
the field and the fluid keep a constant ratio along the evolution.

\bigskip

Another interesting generalization of case 1 arises from a nonvanishing
constant $C$, which yields a family of exponential-like potentials. First we
note that $C\ge 0$ for a real solution. Then, using (\ref{16}),
(\ref{13}) and (\ref{10}) we find the potential in parametric form

$$
\Delta \phi(a) =\frac{\sqrt{6}}s\left\{ \ln a^s+2\left( \frac{6-s}6\right)
^{1/2}
\ln \left[ \left( a^s+c_1\right) ^{1/2}+\left( a^s+c_2\right) ^{1/2}\right]
\right. -
$$
\begin{equation} \label{17a}
\left.
2\ln \left[\left(\frac{6-s}6\right)^{1/2}\left(
a^s+c_1\right)^{1/2}+\left(a^s+c_2\right)^{1/2} \right]
\right \}
\end{equation}

\begin{equation} \label{17b}
V(a)=B a^{s- 6}
\end{equation}

\noindent
where $c_1=sC/((6-s)B)$ and $c_2=sC/(6B)$.
The leading behavior of this potential is (\ref{181}) with
$\sigma=-(6-s)^{1/2}$ for
$\phi\to\infty$, while $\sigma=-(6-n)/\sqrt{6}$ for $\phi\to -\infty$.

When $C> 0$ we find that

\begin{equation} \label{190}
\Delta t=\frac 13\left( \frac s{6Bc_1}\right)^{1/2}a^3\,{}_2F_1
\left( \frac 12,\frac 3s,1+\frac 3s,-\frac{a^s}{c_1}\right)
\end{equation}

\noindent
where ${}_2 F_1$ is the hypergeometric function \cite{Abr}.
These solutions also begin at a singularity, this time like
$a\sim\Delta t^{1/3}$ for $\Delta t\to 0$, while their
asymptotic behavior is like case 1 for $t\to\infty$. Though these solutions
have some initial conditions fixed, we note that they are asymptotically
stable.

 A common framework to solve equations (\ref{6}), (\ref{7}) in discussions
of inflation is the ''slow-roll'' approximation. However, as it has
not been required by our procedure of integration, we can use it to
investigate the limitations imposed by this approximation. Following Copeland
et al. \cite{Cop}, and using equations (\ref{10}) (\ref{11}) we calculate the
slow-roll
parameters

\begin{equation}
\label{16a}
\epsilon \equiv {\frac{{\dot \phi }^2/2}{V+{\dot \phi }^2/2}}=1-{
\frac F{C+6\int da F/a}}
=\left(\frac{6-s}6\right)\frac{a^s+c_1}{a^s+c_2}
\end{equation}

\begin{equation}
\label{16b}
\eta \equiv {\frac{\ddot \phi }{H\dot \phi }}={\frac{
12F-aF^{\prime }-36\int da F/a-6C}{2\left( C+6\int da F/a-F\right) }}
=\frac{\left( s-6\right) a^s-6c_1}{2\left( a^s+c_1\right) }
\end{equation}

The parameter $\epsilon $ measures the relative contribution of the field's
kinetic energy to its total energy density and $\eta $ measures the ratio of
the field's acceleration relative to the friction term. The slow-roll
approximation is valid when $\mid \epsilon \mid \ll 1$ and $\mid \eta \mid \ll
1$. These requirements impose constrains on the form of the potential and the
value of
the initial conditions.

In this case, for $C=0$, the slow-roll constrain reads $6-s\ll 1$, that is,
this approximation is valid only when the potential has a very mild slope and
$\lambda\gg 1$. Moreover, for $C>0$, the evolution near the singularity is
dominated by the kinetic energy. Thus, in general, these solutions cannot be
obtained under the slow-roll approximation.

\subsection{Cosh and cosh-like potentials}

We consider now potentials of the form

\begin{equation} \label{199}
V(\phi)=V_0 \left[\cosh\left(\sigma\Delta\phi\right) \right]^{q}
\end{equation}

\noindent
with $V_0>0$. This potential has a minimum at the origin when $q>0$ or a
maximum when $q<0$ with a value $V_0$.  For large $|\phi|$ it grows or
vanishes exponentially. The case $q<0$ appears when reconstructing the
potential in the assumption of small amplitude of the tensor gravitational
wave spectrum \cite{Cop}. For $q>0$ we have the stable solution $\phi=\phi_0$
and $a(t)$ given by (\ref{150}) with $\Lambda_{eff}=V_0$. We consider bellow
other solutions for this potential.

\bigskip\noindent
Case 1. $F(a) = B a^{s}\left( b + a^{s}\right)^{n}$, where $B>0$, $b> 0$, $s$
and $n$ are constants such that $s(n+1)=6$, $s\neq 0$, $s\neq 6$, and
$\rho_{0}=k=\Lambda =C=0$.

$$
V_0=B\qquad
\sigma=\frac{s}{2\sqrt{6}}\qquad
q=2\left(\frac{6}{s}-1\right)
$$

\noindent
We obtain the time evolution for arbitrary $s$ in implicit form:

\begin{equation}
\Delta t=
\frac{a^3}{\sqrt{3 B} b^{3/s}}\,{}_2F_1\left(\frac{3}{s},\frac{3}{s},
1+\frac{3}{s},-\frac{a^s}{b}\right)
\label{23}
\end{equation}

\begin{equation} \label{23b}
\Delta\phi=-\frac{2\sqrt6}{s} {\rm arccoth}\left(1+\frac{a^s}{b}\right)^{1/2}
\end{equation}

\noindent
 It is suitable for
further analysis to consider explicit expressions for the leading behaviour of
$a(t)$ and $\phi(t)$ in the limit of small and large $a^s/b$

\begin{equation}
a(t)\simeq (\sqrt{3B}b^{3/s}\Delta t)^{1/3},\qquad a^s \ll b
\label{24a}
\end{equation}

\begin{equation}
a(t)\simeq \exp \left[\left(\frac{B}{3}\right)^{1/2}\Delta t\right]
, \qquad a^s \gg b
\label{24b}
\end{equation}

\begin{equation}
\Delta \phi \simeq \left( \frac 23\right) ^{1/2}\ln \Delta t,\qquad\qquad
a^s\ll b
\label{25a}
\end{equation}

\begin{equation}
\Delta \phi \simeq -\frac{2\sqrt{6b}}s\exp \left[ -\frac s2\left( \frac{B}
3\right) ^{1/2}\Delta t\right] ,\qquad a^s\gg b
\label{25b}
\end{equation}

The calculation of the slow-roll parameters

\begin{equation}
\epsilon=\frac{b}{b+a^s}
\label{eps}
\end{equation}

\begin{equation}
\eta=-\frac{1}{2}\left(\frac{6b+sa^s}{b+a^s}\right)
\label{eta}
\end{equation}

\noindent
shows that the slow-roll approximation is valid only for $a^s\gg b$ if
$|s|\ll 1$. So these are new solutions which cannot be obtained under such an
approximation.

We may describe the behavior of these solutions as follows. When $0<s<6$,
$a(t)$ begins at a singularity with the behavior (\ref{24a}) for $\Delta
t\rightarrow 0$ and is asymptotically de Sitter like (\ref{24b}) for $\Delta
t\rightarrow \infty$. The evolution of the field $\phi(t)$ starts dominated by
a diverging kinetic energy and settles down at the minimum of the potential
for large times driving the exponential expansion. When
$s<0$ the evolution begins like (\ref{24b}) in the remote past and then
deflates into the asymptotic Friedmann behavior (\ref{24a}) in the far future.
The field begins rolling down very slowly from the maximum of the potential.
Then it gains kinetic energy that decays much more slowly than the potential
energy. As it becomes dominant, it drives the deflationary era and makes the
field grow without bound. We provide in this way an exact realization of the
deflationary universe scenario proposed recently by Spokoiny \cite{Spo}.
We note also that for either sign of $s$, these solutions are stable. Finally,
when $s>6$, the scale factor behaves the same as the case $0<s<6$, but the
field begins its evolution dominated by the kinetic energy and approaches very
slowly to the maximum of the potential for large times. This is an unstable
solution.

\bigskip\noindent
Case 2. $F(a)=B a^s$, $\Lambda=C=0$, $s=6-3\gamma$, $0<\gamma<2$, $k=1$.

$$
V_0=B\left[ \frac 13\left( \frac {2B}{2-\gamma }+\rho _0\right) \right]
^{\frac{3\gamma }{2-3\gamma }}
$$

$$
\sigma =\frac{3\gamma -2}2\left\{ \frac 1{6\gamma }\left[ 2+\frac{\rho
_0}B\left( 2-\gamma \right) \right] \right\} ^{1/2}\qquad
q=\frac{6\gamma }{3\gamma -2}
$$

\noindent
The time evolution is given by

\begin{equation} \label{27}
\Delta t=\frac{2}{3\gamma }\left(\frac{\omega}{B}\right)^{1/2}a^{3\gamma
/2}\,{}_2F_1\left( \frac
12,\frac{3\gamma }{6\gamma -4},\frac 32-\frac 1{2-3\gamma },\omega a^{3\gamma
-2}\right)
\end{equation}

\begin{equation} \label{28}
\Delta\phi=-\frac{1}{\sigma}{\rm arctanh} \left(1-\omega
a^{3\gamma-2}\right)^{1/2}
\end{equation}

\noindent
where

$$
\omega=3\left(\frac{2}{2-\gamma}+\frac{\rho_0}{B}\right)^{-1}
$$

\noindent
For $0<\gamma<2/3$ ($q<0$), the scale factor has a bounce and its asymptotic
behavior for large times is $t^{2/(3\gamma)}$. For $2/3<\gamma<2$ ($q>0$), the
evolution starts from a singularity with the leading behavior $\Delta
t^{2/(3\gamma)}$, it reaches a maximum and then collapses again. We note that
for
the physically interesting case of radiation $\gamma=4/3$, the solution takes
a simple closed form:

\begin{equation} \label{29}
a(t)=\left(B+\frac{\rho_0}{3}-\Delta t^2\right)^{1/2}
\end{equation}

\bigskip\noindent
Case 3. $F(a)=B a^s$, $\rho_0=\Lambda=C=0$, $0<s<6$, $s\neq 4$, $k=1$.

$$
V_0=B\left(\frac{s}{2B}\right)^{\frac{6-s}{4-s}}\qquad
\sigma=\frac{4-s}{2\sqrt{6-s}}\qquad
q=\frac{2(6-s)}{4-s}
$$

\noindent
The time evolution is given by

\begin{equation} \label{30}
\Delta t=\left(\frac{2s}{B}\right)^{1/2} \frac{a^{3-s/2}}{6-s}\,{}_2F_1
\left(\frac{1}{2},\frac{s-6}{2(s-4)},\frac{3s-14}{2(s-4)},
\frac{sa^{4-s}}{2B}\right)
\end{equation}

\begin{equation} \label{31}
\Delta\phi=-\frac{1}{\sigma} {\rm arctanh} \left(1-\frac{s
a^{4-s}}{2B}\right)^{1/2}
\end{equation}

\noindent
For $4<s<6$ ($q<0$), the scale factor has a bounce and its asymptotic
behavior for large times is $t^{2/(6-s)}$. For $0<s<4$ ($q>0$), the
scale factor starts from a singularity with the leading behavior $\Delta
t^{2/(6-s)}$, it reaches a maximum and then collapses again.

\bigskip
We show here  another potential that have qualitatively the form
(\ref{199}):

\begin{equation} \label{310}
V(\phi)=B\left[\frac{\epsilon\exp(\sigma\Delta\phi)}
{\frac{(2+3\gamma)}{24}\left(1+3\epsilon\frac{k}{B}\exp(\sigma\Delta\phi)
\right)^{2}-\frac{\rho_0}{B}\exp(2\sigma\Delta\phi)}\right]^
{\frac{3\gamma-4}{2-3\gamma}}
\end{equation}

\noindent
where $\epsilon=\pm 1$ and

\begin{equation} \label{3101}
\sigma=\frac{2-3\gamma}{\sqrt{(4-3\gamma)}}
\end{equation}

 They arise in the case that $F(a)=Ba^s$,
$\Lambda=C=0$, $s=2+3\gamma$ and $k=1$. There is  a critical energy density
coefficient
$\rho_c=3(2+3\gamma)/(8B)$ and for $\rho_0<\rho_c$ we find a cosh-like
potential with $q>0$ when either $\gamma<2/3$
($\epsilon=1$), $2/3<\gamma<4/3$ ($\epsilon=-1$) or
$2/3<\gamma<4/3$ ($\epsilon=1$), this time $q<0$.
The scale factor starts from a singularity with the
leading behavior $\Delta t^{2/(4-3\gamma)}$, it reaches a maximum and then
recollapses again.

We see from our examples that we cannot discard a priori the influence of a
curvature term in the determination of the evolution, since a potential that
leads in the flat case to an asymptotically expanding scale factor yields
recollapsing solutions when $k=1$.

\subsection{Sinh and sinh-like potentials}

We consider potentials of the form

\begin{equation} \label{311}
V(\phi)=V_0 \left|\sinh(\sigma \Delta\phi)\right|^{q}
\end{equation}

\noindent
with $V_0>0$. This potential has a vanishing minimum at the origin when $q>0$
or diverges when $q<0$. For large $|\phi|$ it grows or vanishes exponentially.
This kind of potentials satisfy, for suitable values of $q>0$, the
requirements usually imposed in the chaotic inflationary scenario \cite{Lin}.
We study several exact solutions for this potential.

\bigskip\noindent
Case 1. $F(a)=B a^s$, $\Lambda=C=0$, $s=6-3\gamma$, $0<\gamma<2$, $k=-1$.

$$
V_0=B\left[ \frac 13\left( \frac {2B}{2-\gamma }+\rho _0\right) \right]
^{\frac{3\gamma }{2-3\gamma }}
$$

$$
\sigma =\frac{3\gamma -2}2\left\{ \frac 1{6\gamma }\left[ 2+\frac{\rho
_0}B\left( 2-\gamma \right) \right] \right\} ^{1/2}\qquad
q=\frac{6\gamma }{3\gamma -2}
$$

\noindent
The time evolution is given by

\begin{equation} \label{32}
\Delta t=a\,{}_2F_1\left(\frac{1}{2},\frac{1}{2-3\gamma},
1+\frac{1}{2-3\gamma},-\frac{a^{2-3\gamma}}{\omega}\right)
\end{equation}

\begin{equation} \label{33}
\Delta\phi=-\frac{1}{\sigma}{\rm arccoth} \left(1+\omega
a^{3\gamma-2}\right)^{1/2}
\end{equation}

\noindent
The evolution of the scale factor for these models starts from a singularity
as $\Delta t^{\lambda_1}$ for $\Delta t\to 0$ and grows monotonically with
asymptotic behavior $t^{\lambda_2}$ for $t\to\infty$. In this case, for
$0<\gamma<2/3$, $\lambda_1=1$ and $\lambda_2=2/(3\gamma)>1$, while for
$2/3<\gamma<2$, $\lambda_1=2/(3\gamma)<1$ and $\lambda_2=1$.

\bigskip\noindent
Case 2. $F(a)=Ba^s$, $k=\Lambda=C=0$, $0<s<6$, $6-3\gamma-s\neq 0$.

$$
V_0=B\left(\frac{s\rho_0}{6B}\right)^{\frac{6-s}{6-3\gamma-s}}\qquad
\sigma=\frac{6-3\gamma-s}{2\sqrt{6-s}}\qquad
q=\frac{2(6-s)}{6-3\gamma-s}
$$

The time evolution is given by

\begin{equation} \label{34}
\Delta t=\left(\frac{2s}{B}\right)^{1/2} \frac{a^{3-s/2}}{6-s}\,
{}_2F_1\left(\frac{1}{2},\frac{6-s}{2(6-3\gamma-s)},
\frac{3(6-2\gamma-s)}{2(6-3\gamma-s)},-\frac{\rho_0s}{6B}a^{6-3\gamma-s}\right)
\end{equation}

\begin{equation} \label{35}
\Delta\phi=-\frac{1}{\sigma}{\rm arccoth}
\left(1+\frac{s\rho_0}{6B}a^{6-3\gamma-s}\right)^{1/2}
\end{equation}

\noindent
For $6-3\gamma-s>0$,  $q>0$, $\lambda_1=2/(6-s)$ and $\lambda_2=2/(3\gamma)$.
Then,
$\lambda_1<\lambda_2$ but this is not an inflationary solution unless
$\gamma<2/3$. For
$6-3\gamma-s<0$, $q<0$, $\lambda_1=2/(3\gamma)$ and $\lambda_2=2/(6-s)$. So
$\lambda_1<\lambda_2$, and
this model may describe the passage from a primordial radiation-dominated era
to a power-law inflationary stage.

\bigskip\noindent
Case 3. $F(a)=Ba^s$, $\Lambda=C=0$, $s=4$, $k=0,-1$ or $k=1$ if $B>2$.

$$
V_0=B\left[\frac{3}{\rho_0}\left(\frac{B}{2}-k\right)\right]^
{\frac{2}{3\gamma-2}}\qquad
\sigma=\frac{3\gamma-2}{2}\left(\frac{1}{2}-\frac{k}{B}\right)^{1/2}\qquad
q=\frac{4}{2-3\gamma}
$$

\noindent
The time evolution is given by

\begin{equation} \label{36}
\Delta t=\frac{a}{\left(\frac{B}{2}-k\right)^{1/2}}\,
{}_2F_1\left(\frac{1}{2},\frac{1}{2-3\gamma},\frac{3(1-\gamma)}{2-3\gamma},
-\omega a^{2-3\gamma}\right)
\end{equation}

\begin{equation} \label{37}
\Delta\phi=-\frac{1}{\sigma}{\rm arccoth} \left(1+\omega
a^{2-3\gamma}\right)^{1/2}
\end{equation}

\noindent
For $\gamma<2/3$ $q>0$ and we find $\lambda_1=1$ and $\lambda_2=2/(3\gamma)>1$.
For
$\gamma>2/3$ $q<0$, $\lambda_1=2/(3\gamma)$ and $\lambda_2=1$.

\bigskip\noindent
Case 4. $F(a)=Ba^s$, $\rho_0=\Lambda=C=0$, $0<s<6$, $s\neq 4$, $k=-1$.

$$
V_0=B\left(\frac{s}{2B}\right)^{\frac{6-s}{4-s}}\qquad
\sigma=\frac{4-s}{2\sqrt{6-B}}\qquad
q=\frac{2(6-s)}{4-s}
$$

\noindent
The time evolution is given by

\begin{equation} \label{38}
\Delta t=a \,{}_2F_1\left(\frac{1}{2},\frac{1}{s-4},\frac{s-3}{s-4},
-\frac{2B}{s}a^{s-4}\right)
\end{equation}

\begin{equation} \label{39}
\Delta \phi=-\frac{1}{\sigma} {\rm arccoth}
\left(1+\frac{s}{2B}a^{4-s}\right)^{\frac{1}{2}}
\end{equation}

\noindent
For $0<s<4$ $q>0$, $\lambda_1=2/(6-s)<1$ and $\lambda_2=1$. For $4<s<6$, $q<0$,
$\lambda_1=1$
and $\lambda_2=2/(6-s)>1$, so that the evolution is inflationary.

\bigskip
The expression (\ref{310}) ($\epsilon=1$) also yields potentials of the sinh
form, with
$q>0$ for $\rho_0>\rho_c$. They occur when $k=1$ and $\gamma<2/3$, or $k=-1$
and $0<\gamma<4/3$, $\gamma\neq 2/3$. The scale factor starts from a
singularity and grows monotonically. For $0<\gamma<2/3$
$\lambda_1=2/(4-3\gamma)<1$ and $\lambda_2=2/(3\gamma)>1$, while for
$2/3<\gamma<4/3$ $\lambda_1=2/(3\gamma)<1$ and $\lambda_2=2/(4-3\gamma)>1$. In
either case there are particle horizons and the evolution is inflationary.

We note that all the solutions presented for potential (\ref{311}) are
asymptotically stable. Many of them show the influence of the fluid source on
their behavior, either near the singularity or for large times.

\section{Conclusions}

We  have shown that the Einstein equations with a self-interacting
minimally coupled scalar field, a perfect fluid source and
cosmological constant can be reduced to quadratures in the Robertson-Walker
metric . For that, the scale factor is taken as the independent
variable and the potential is expressed in terms of it. This is fully
justified as long as $a(t)$ and $\phi(t)$ are monotonic functions. For this
class of cosmological models we have reobtained in a unified manner many exact
solutions presented by other authors, and found several new exact solutions
for some simple interesting potentials. We have investigated their behavior in
detail, and shown that most of the solutions that do not recollapse in a
finite time are asymptotically and structurally stable.

For a given potential, the influence of a fluid source
or the spatial curvature may dominate the evolution of the early universe. In
effect, an asymptotically inflationary behavior cannot occur when the energy
density of the scalar field decays faster than the energy density of a fluid
with $\gamma>2/3$. We have shown asymptotically stable solutions such that
both densities keep a constant ratio, or one of them dominates for large
times. All the same, a negative spatial curvature may lead to a linearly
expanding scale factor. On the other hand, a positive curvature has frequently
as a consequence bouncing solutions or solutions that undergo recollapse to a
second singularity. In such cases, the energy density of the scalar field may
increase, instead of rolling down the potential.

We have studied the limitations imposed by the slow-roll approximation and
most of our solutions cannot be obtained by means of this
approximation. Calculations of the primordial spectrum of perturbations, based
on recent measurements of the cosmological background radiation show that
tensorial perturbations may have played an important role in the formation of
cosmic structures. However, the amount of gravitational perturbations
predicted by means of the slow-roll approximation is very small \cite{Lit}.
Thus exact solutions which lay outside the slow-roll regime may lead to an
improved understanding of the evolution of the early universe.


\newpage
\begin{thebibliography}{9}

\bibitem{Gut}
Guth A H 1981
{\it Phys. Rev.\/} D {\bf 23} 347

\bibitem{Alb}
Albrecht A and Steinhard P J  1982
{\it Phys. Rev. Lett.\/} {\bf 48} 1220

\bibitem{Lin82}
Linde A D 1982
{\it Phys.  Lett.\/} {\bf 108B} 389

\bibitem{Lin}
Linde A D 1990
{\it Particle Physics and Inflationary Cosmology\/} (Chur: Harwood)

\bibitem{Kol}
Kolb E W and Turner M S 1990
{\it The Early Universe\/} (Redwood City: Addison Wesley)

\bibitem{Rat}
Ratra B and Peebles P J E    1988
{\it Phys. Rev. \/} {\bf 37}  3406

\bibitem{Ell}
Ellis G F R and Madsen M S 1991
{\it Class. Quantum Grav.\/} {\bf 8} 667

\bibitem{Cop}
Copeland E J, Kolb E~W, Liddle A~R and Lidsey J~E 1993
{\it Phys. Rev. \/} D {\bf 48}  2529

\bibitem{Lid94}
Liddle A~R and Turner M S 1994
{\it Phys. Rev. \/} D {\bf 50} 758

\bibitem{Luc}
Lucchin F and Matarrese S 1985
{\it Phys. Rev.\/} D {\bf 32} 1316

\bibitem{Bur}
Burd A B and Barrow J D 1988
{\it Nucl. Phys. \/} {\bf 308B} 929

\bibitem{Rit1}
de Ritis R, Platania G, Scudellaro P and Stornaiolo C 1990
{\it Gen. Rel. Grav. \/} {\bf 22} 97

\bibitem{Rit2}
de Ritis R, Marmo G, Platania G, Rubano C, Scudellaro P and Stornaiolo C 1990
{\it Phys. Rev. \/} D {\bf 42} 1091, {\it Phys. Lett. \/} {\bf 149A} 79

\bibitem{Mus}
Muslimov A G 1990
{\it Class. Quantum Grav.\/} {\bf 7} 231

\bibitem{Sal}
Salopek D S and Bond J R 1990
{\it Phys. Rev. \/} D {\bf 42} 3962

\bibitem{Bar90}
Barrow J D 1990
{\it Phys. Lett. \/} {\bf 235B} 40

\bibitem{Chi}
Chimento L P, Cossarini A E and Jakubi A S
"Exact self-interacting scalar field cosmologies"
 Proceedings of {\it Aspect of General Relativity and
Mathematical Physics, Mexico City, Mexico, 1993\/}

\bibitem{Bar94}
Barrow J D 1994
{\it Phys. Rev.\/} D {\bf 49} 3055

\bibitem{Mie}
Schunck F E and Mielke E W 1994
{\it Phys. Rev.\/} D {\bf 50} 4794

\bibitem{Hal}
Halliwell J J 1987
{\it Phys. Lett. \/} {\bf 185B} 341

\bibitem{Oze}
\"Ozer M and Taha M O 1992
{\it Phys. Rev.\/} D {\bf 45} 997

\bibitem{Eas}
Easther R 1993
{\it Class. Quantum Grav.\/} {\bf 10} 2203

\bibitem{Bar93}
Barrow J D 1993
{\it Class. Quantum Grav.\/} {\bf 10} 279

\bibitem{Cos}
Chimento L P and Cossarini A E 1994
{\it Class. Quantum Grav. \/} {\bf 11} 1177


\bibitem{Kra}
Krasovskii  N N 1963
{\it Stability of Motion \/} (Stanford University Press: Stanford)

\bibitem{Har}
Harrison E R 1967
{\it Mon. Not. R. Ast. Soc.\/} {\bf 137} 69


\bibitem{Ell91b}
Ellis G F R, Skea J E F and Tavakol R K 1991
{\it Europhys. Lett. \/} {\bf 16} 767

\bibitem{Lid93}
Lidsey J E 1993
{\it Gen. Rel. Grav. \/} {\bf 25} 399

\bibitem{Abr}
Abramowitz M and Stegun I A eds 1965
{\it Handbook of Mathematical Functions \/} (Dover: New York)

\bibitem{Spo}
Spokoiny B 1993
{\it Phys. Lett. \/} {\bf 315B} 40

\bibitem{Lit}
Liddle A R and Lyth D H 1993
{\it Phys. Rep. \/} {\bf 231} 1

\end{thebibliography}
\end{document}